\documentclass[aps, pra, reprint, superscriptaddress]{revtex4-2}

\usepackage{helvet}
\usepackage[utf8]{inputenc}
\usepackage{amsmath}
\usepackage{amssymb}
\usepackage{hyperref}
\usepackage{graphicx}
\usepackage{epstopdf}
\usepackage{dcolumn}
\usepackage{soul}
\usepackage{bm}
\usepackage{soul}
\usepackage{xspace}
\usepackage{verbatim}
\usepackage{color}
\usepackage{xcolor}

\hypersetup{colorlinks=true,linkcolor=blue,citecolor=blue,filecolor=blue,urlcolor=blue,breaklinks=true}

\begin{document}

\title{Persisting quantum effects in the anisotropic Rabi model at thermal equilibrium}

\author{He-Guang Xu}
\affiliation{School of Physics, Dalian University of Technology, 116024 Dalian,  China}

\author{V. Montenegro}
\email{vmontenegro@uestc.edu.cn}
\affiliation{Institute of Fundamental and Frontier Sciences, University of
	Electronic Science and Technology of China, Chengdu 610051, China}

\author{Gao Xianlong }
\email{gaoxl@zjnu.edu.cn}
\affiliation{Department of Physics, Zhejiang Normal University, Jinhua 321004, Zhejiang, P. R. China}

\author{Jiasen Jin}
\email{jsjin@dlut.edu.cn}
\affiliation{School of Physics, Dalian University of Technology, 116024 Dalian,  China}

\author{G. D. de Moraes Neto}
\email{gdmneto@zjnu.edu.cn}
\affiliation{Department of Physics, Zhejiang Normal University, Jinhua 321004, Zhejiang, P. R. China}

%\date{\today}

\begin{abstract}
Quantum correlations and nonclassical states are at the heart of emerging quantum technologies. Efforts to produce long-lived states of such quantum resources are a subject of tireless pursuit. Among several platforms useful for quantum technology, the mature quantum system of light-matter interactions offers unprecedented advantages due to current on-chip nanofabrication, efficient quantum control of its constituents, and its wide range of operational regimes. Recently, a continuous transition between the Jaynes-Cummings model and the Rabi model has been proposed by exploiting anisotropies in their light-matter interactions, known as the anisotropic quantum Rabi model. In this work, we study the long-lived quantum correlations and nonclassical states generated in the anisotropic Rabi model and how these indeed persist even at thermal equilibrium. To achieve this, we thoroughly analyze several quantumness quantifiers, where the long-lived quantum state is obtained from a dressed master equation that is valid for all coupling regimes and with the steady state ensured to be the canonical Gibbs state. Furthermore, we demonstrate a stark distinction between virtual excitations produced beyond the strong coupling regime and the quantumness quantifiers once the light-matter interaction has been switched off. This raises the key question about the nature of the equilibrium quantum features generated in the anisotropic quantum Rabi model and paves the way for future experimental investigations, without the need for challenging ground-state cooling.

%Quantum resources have played a crucial role in the development of cutting-edge technologies as a critical resource. This work investigates the harvesting potential of quantum resources in the open anisotropic Rabi model. We employ a quantum Markovian master equation to describe the evolution of the system that is valid for all coupling regimes and verify that the steady state is a Gibbs thermal equilibrium state. Furthermore, through a set of complementary measurements that witness different quantum aspects, we characterize the quantumness on the finite-temperature-coupling strength phase diagram.
\end{abstract}

%keywords
%\pacs{42.50.Ar, 03.65.Yz, 42.50.Pq}

\maketitle

\section{Introduction}
The resource theory of quantum information~\cite{chitambar2019quantum,streltsov2017colloquium} demonstrates that quantum correlations found in nonclassical states are of utmost importance for performing quantum processing tasks~\cite{acin2018quantum,nielsen2010quantum}. From a theoretical perspective, nonclassical states have proven highly valuable in investigating decoherence~\cite{zurek2003decoherence}, witnessing the potential quantum nature of gravity~\cite{bose2017spin,marletto2017gravitationally}, studying the quantum-to-classical transition~\cite{modi2012classical}, enabling remote quantum control of the weak value amplification~\cite{Coto_2017}, and serving as a powerful tool for witnessing quantum phase transitions in critical systems~\cite{amico2008entanglement}. From a practical standpoint, quantum correlations have demonstrated their pivotal role as a fundamental element in emerging quantum technologies~\cite{boixo2018characterizing,preskill2018quantum}, including secure quantum communication~\cite{gisin2007quantum,QKD2009Review}, quantum sensing~\cite{degen2017quantum,Giovannetti2011,Giovannetti2006,Giovannetti2004,montenegro2023quantum,montenegro2022probing,montenegro2020mechanical}, and quantum simulation~\cite{Simulation2014Review}. Nonetheless, the intrinsic sources of noise and decoherence inherent in quantum dynamics render the production of long-lived quantum states necessary for quantum technologies~\cite{suter2016colloquium,eremeev2012thermally}. Efforts to advance efficient techniques for generating and protecting nonclassical states against quantum noise~\cite{duan1997preserving} encompass approaches such as decoherence-free subspaces~\cite{mundarain2007decoherence,lidar1998decoherence,lidar2000protecting}, dynamical decoupling~\cite{viola2005random,celeri2008switching}, and reservoir engineering schemes~\cite{reiter2016scalable,de2017steady,de2014steady}. Therefore, proposing schemes that facilitate the generation of long-lived highly correlated nonclassical states is one of the primary tasks within the quantum information field. 

Quantum systems undergoing light-matter interactions stand as diverse platforms for generating nonclassical states and executing quantum information tasks~\cite{blais2020quantum,wendin2017quantum}. The recently achieved ultrastrong-coupling regime (USC)~\cite{anappara2009signatures,niemczyk2010circuit,forn2019ultrastrong} and deep strong coupling regime (DSC)~\cite{yoshihara2017superconducting,yoshihara2018inversion} of light-matter interactions have advanced quantum correlation generation beyond the strong coupling regime (SC)~\cite{meschede1985one,thompson1992observation,weisbuch1992observation,lodahl2015interfacing,gu2017microwave}, i.e. when the light-matter coupling strength exceeds both the decay rates and the natural frequencies of the systems. Notably, a clear distinction between USC/DSC and the SC regime is the emergence of nonclassicality in the ground state of the system, manifested through squeezing and entanglement~\cite{ashhab2010qubit,shen2014ground}. Moreover, within the USC regime, thermal photons can exhibit anti-bunching behavior~\cite{ridolfo2013nonclassical}, and emissions from the matter subsystem (e.g., a two-level atom) can display photon bunching~\cite{garziano2017cavity}, whereas the behavior is reversed in the SC regime.

The cornerstone model describing light-matter interaction is composed of a two-level atom (qubit) coupled to a quantized single-mode cavity field through dipole interaction, known as the quantum Rabi model (QRM)~\cite{rabi1936process,rabi1937space,braak2016semi}. The QRM has been the subject of extensive theoretical work in quantum optics~\cite{scully1997quantum}, for the generation of quantum entanglement~\cite{chen2010entanglement}, as an Otto quantum engine in quantum thermodynamics~\cite{altintas2015rabi}, to exhibit the emergence of quantum phase transitions~\cite{hwang2015quantum,hwang2016recurrent,liu2017universal}, and motivated the development of a novel operational criterion for integrability~\cite{braak2011integrability,chen2012exact}. Interestingly, the QRM transforms into the Jaynes-Cummings model (JCM)~\cite{jaynes1963comparison} in situations where the natural frequencies of the system significantly surpass the bare light-matter coupling strength. In this scenario, one can invoke the rotating-wave approximation (RWA), permitting the retention of the rotating-wave (RW) terms while omitting the counter-rotating-wave (CRW) terms. However, this approximation becomes invalid once the system transit towards the USC regime~\cite{chen2012exact,de2007quantum,peropadre2010switchable,nataf2010vacuum,casanova2010deep}.

This motivates the introduction of a continuous transition between the JCM and the QRM by varying the RW and CRW coupling strengths, known as the anisotropic quantum Rabi model (AQRM). The AQRM has been explored in both closed systems~\cite{yang2017ultrastrong,shen2017quantum} and open systems~\cite{joshi2016quantum,ye2023implication}, with novel quantum technological schemes such as criticality-enhanced quantum sensing~\cite{zhu2023criticality} and squeezed vacuum state laser~\cite{de2022squeezed}. Notably, an exact solution for the AQRM with a biased term has been recently derived~\cite{xie2014anisotropic}, and certain analytical solutions have been found via transcendental function extension~\cite{xie2014anisotropic,chen2021multiple} and transformation techniques~\cite{zhang2015analytical,zhang2016generalized}. Experimental realizations of the AQRM have been achieved in two-dimensional quantum wells~\cite{wang2016energy}, cavity quantum electrodynamics~\cite{grimsmo2013cavity}, and superconducting circuits~\cite{xie2014anisotropic}. The flexibility to adjust the RW and CRW coupling strengths renders the AQRM to be highly versatile. Indeed, this grants the ability to explore a plethora of coupling regimes, thereby enabling the creation of highly-correlated nonclassical states. A pivotal question arises: can we generate long-lived highly-correlated nonclassical states that persist even in the presence of a reservoir at finite temperature? This question is of paramount importance for current experimental platforms.

In this work, we demonstrate that long-lived highly-correlated nonclassical states of the AQRM can indeed persist even in the presence of a reservoir at finite temperature. To support our theoretical findings, we use a quantum Markovian master equation to describe the evolution of the AQRM. This approach remains applicable across all coupling regimes, including degenerate points in the energy spectrum, i.e. without resorting to secular approximation. Consequently, such a quantum open system genuinely approaches a thermal Gibbs state as its steady state. Furthermore, through a series of complementary analyses, we characterize the \textit{quantumness} of the steady state in the finite-temperature-coupling strength phase diagram, showing that highly-correlated nonclassical states can be achieved across a wide parameter region.

The rest of the paper is structured as follows: In Sec.~\ref{The model and methods} we present the open system description of our model. In Sec.~\ref{quantifiers} we briefly defined the measures and witnesses of nonclassicality thorughout our work. In Sec.~\ref{results}, we outline our results, including the effects of temperature in the system. Finally, we present our concluding remarks in Sec.~\ref{conclusion}.

\section{The model and methods}\label{The model and methods}
The Hamiltonian of the AQRM~\cite{xie2014anisotropic} is ($\hbar = 1$)
\begin{equation}
	\hat{H}=\omega\hat{a}^{\dagger}\hat{a}+\frac{\Delta}{2}\hat{\sigma}_{z}+\lambda_{1}(\hat{a}\hat{\sigma}^{+}+\hat{a}^{\dagger}\hat{\sigma}^{-})+\lambda_{2}(\hat{a}^{\dagger}\hat{\sigma}^{+}+\hat{a}\hat{\sigma}^{-}).\label{eq:Hamiltonian-aqrm}
\end{equation}
This quantum system couples a two-level system (qubit) with a single-mode bosonic field undergoing both the RW and the CRW interactions terms unevenly. Here, the qubit is described by Pauli matrices $\hat{\sigma}_{x,y,z}, \hat{\sigma}^{\pm}=(\hat{\sigma}_{x}\pm i\hat{\sigma}_{y})/2$ and transition frequency $\Delta$, whereas the boson field is described with frequency $\omega$ and annihilation (creation) operator $\hat{a}$ ($\hat{a}^\dagger$). Finally, $\lambda_1$ ($\lambda_2$) accounts for the light-matter coupling strength for the RW (CRW) interaction, where $\lambda_2=0$ is the JCM limit, $\lambda_1 = 0$ is the anti-Jaynes-Cummings model (AJCM), and $\lambda_1=\lambda_2$ is the isotropic QRM.

One key aspect of the AQRM in Eq.~\eqref{eq:Hamiltonian-aqrm} is that preserves the parity symmetry $\mathbb{Z}_{2}$ as in the isotropic model~\cite{forn2019ultrastrong}. This fact can be evidenced by considering the total number of excitations operator $\hat{n}=\hat{a}^\dagger\hat{a}+\hat{\sigma}^+\hat{\sigma}^-$ and the parity operator $\hat{\pi}=\mathrm{exp}\left(i\pi\hat{n}\right)$. One can readily notice that, as opposed to the JCM, for the AQRM case $[\hat{H},\hat{n}]\neq 0$ and $[\hat{H},\hat{\pi}]=0$. The above has two immediate consequences: (i) even in the ground state, the expected mean number of qubit and boson excitations is non-zero, and (ii) it implies that $\hat{\pi}$ possesses two eigenvalues, $\left\langle\hat{\pi}\right\rangle = \pm 1$, depending on whether the total number of excitations are even or odd. This in turns enable to solve the AQRM analytically by similar techniques used for the isotropic case, for instance, the Bargmann-Fock space of analytic functions and the Bogoliubov operator approach.

\begin{figure*}
\centering \includegraphics[width=0.75\linewidth]{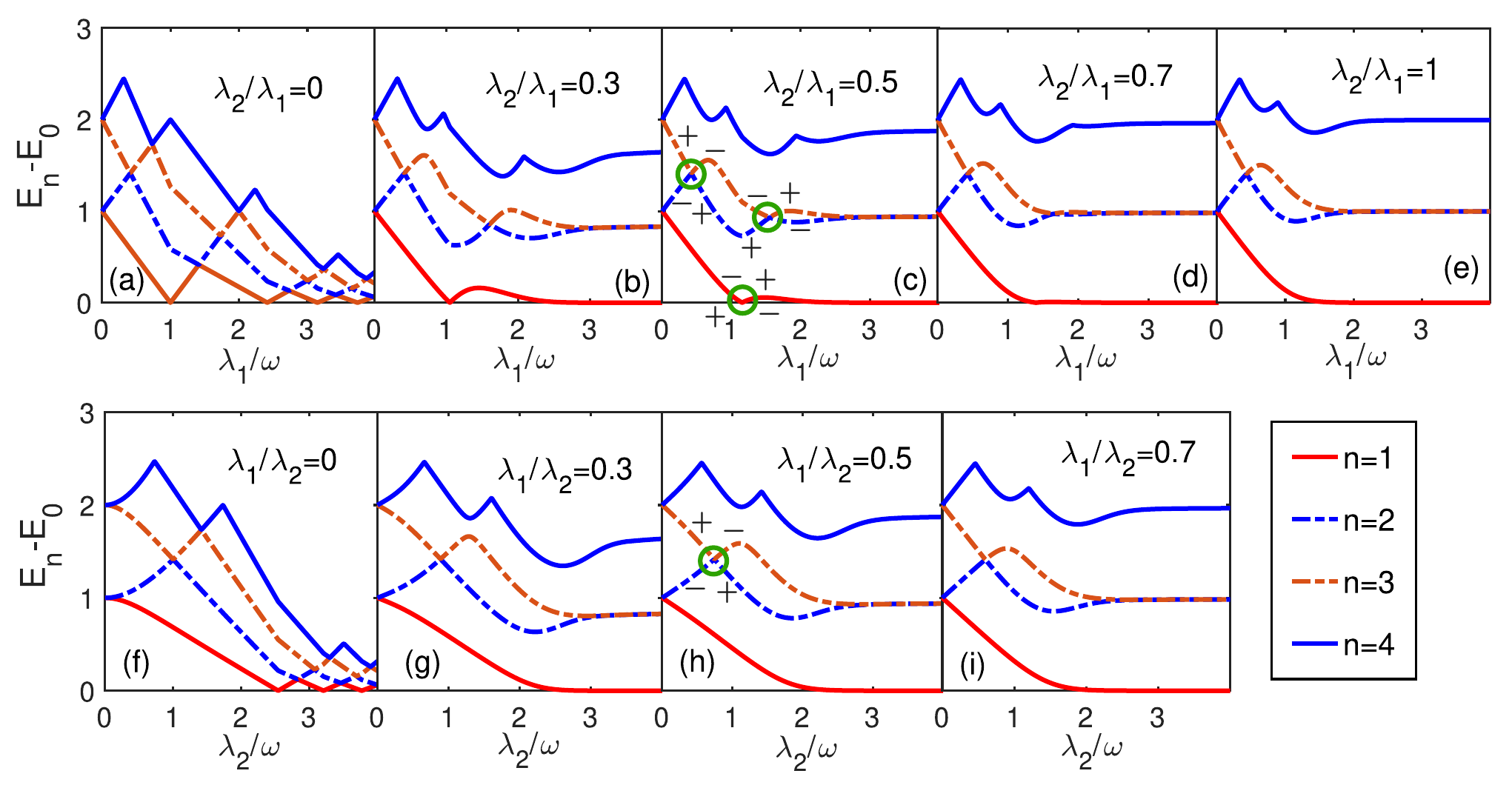}
\caption{Energy level differences are plotted as functions of the light-matter coupling strengths $\lambda_{1}$ ($\lambda_{2}$) for different coupling ratios $\lambda_{2}/\lambda_{1}$ ($\lambda_{1}/\lambda_{2}$). In (c) and (h), the green circles represent the (degenerate) crossing of the energy levels and the plus and minus signs denote the change of parity of the energy levels. Other system parameters are specified as $\Delta=1$ and $\omega=1$.}~\label{fig1}
\end{figure*}

To gain a clearer understanding of the energy levels of the AQRM, in Fig.~\ref{fig1}, we plot the energy spectrum in relation to the ground state energy as a function of the light-matter coupling $\lambda_1$ ($\lambda_2$) for several coupling ratios $\lambda_2/\lambda_1$ ($\lambda_1/\lambda_2$). As seen from the figure, the spectrum exhibits level crossings where the ground state parity changes sign (see green circles), leading the system to undergo infinite first-order quantum phase transitions as the coupling strength increases. In addition to the above degenerate crossing, several other non-degenerate and quasi-degenerate points can be observed. The analysis of the AQRM spectrum, as shown in Fig.~\ref{fig1}, is important for understanding the physics underlying the emergence of thermal correlations, a study that we will address in later sections. 

%==========================================
%==========================================

As stated before, the central objective of our work is to study the quantum correlations and quantum nonclassicality, which remain present even at thermal equilibrium ---the so-called thermal correlations and thermal nonclassicality of the AQRM, respectively. See quantum thermalization and thermal entanglement in the open JCM~\cite{fan2020quantum} and QRM~\cite{liu2023quantum} as particular cases of the AQRM. To do so, we obtain the AQRM steady-state when such a system interacts unavoidably with a reservoir at temperature $T$. It is essential to point out that, even though the standard Born-Markov master equation accurately describes the quantum state within the weak coupling regime $\lambda_i\ll\{\omega, \Delta\}$, it breaks down for stronger coupling regimes such as the USC and DSC regimes~\cite{frisk2019ultrastrong,niemczyk2010circuit,forn2019ultrastrong}. Consequently, the steady-state at thermal equilibrium cannot be fully captured using the standard master equation in those regimes. To ensure that the steady-state is the actual quantum state at thermal equilibrium for an arbitrary set of coupling strengths $\lambda_i$, one needs to switch to a Born-Markov master equation in a dressed picture~\cite{BRE02}, see Refs.~\cite{Scala_2007, montenegro2011creation} for a similar approach in the JC case. In this more suitable representation, the quantum jumps occur between the dressed eigenstates ladder of the full Hamiltonian as opposed to the standard master equation where the upward and downward jumps only consider the free energy of the Hamiltonian.

We emphasize that, the AQRM spectrum (see Fig.~\ref{fig1}) shows non-degenerate and degenerate eigenvalues which are appropriately modeled by the secular approximation in the derivation of the Markovian master equation, but it fails in the quasi-degenerate case. Following the steps of Ref.~\cite{mccauley2020accurate}, we can derive a valid master equation for the entire spectrum of the Hamiltonian Eq.~\eqref{eq:Hamiltonian-aqrm}, as long as we consider weak damping, a high bath cut-off frequency, and a flat spectral density, these are sufficient conditions to ensure positivity and Markovianity. Indeed, the dissipative dynamics undergoes the following dressed master equation
\begin{eqnarray}
	\frac{d}{dt}\hat{\rho}&=&-i[\hat{H},\hat{\rho}]+\sum_{ \substack{u=a,\sigma^{-}\\k<j } }
	\{\Gamma^{jk}_un_u(\Delta_{jk})\mathfrak{D}[|\phi_j{\rangle}{\langle}\phi_k|,\hat{\rho}]\nonumber\\
	&&+\Gamma^{jk}_u[1+n_u(\Delta_{jk})]\mathfrak{D}[|\phi_k{\rangle}{\langle}\phi_j|,\hat{\rho}]\},\label{eq:dressed-me}
\end{eqnarray}
with Lindbladian superoperator term defined as 

\begin{equation}
	\mathfrak{D}[\hat{O},\hat{\rho}]=\frac{1}{2}[2\hat{O}\hat{\rho}\hat{O}^{\dag}-\hat{\rho}\hat{O}^{\dag}\hat{O}-\hat{O}^{\dag}\hat{O}\hat{\rho}].
\end{equation}
In Eq.~\eqref{eq:dressed-me}, $|\phi_k{\rangle}$ is the eigenvector of the AQRM, namely

\begin{equation}
	\hat{H}|\phi_k{\rangle}=E_k|\phi_k{\rangle},
\end{equation}
where $E_k$ is the $k$th eigenenergy associated to the eigenstate $|\phi_k{\rangle}$. We also define the dissipation rates 
\begin{equation}
\Gamma^{jk}_u=\gamma_u(\Delta_{jk})|S^{jk}_u|^2,
\end{equation}
with two explicit contributions, 

\noindent (i) The spectral function:
\begin{equation}
\gamma_u(\Delta_{jk})=2\pi\sum_k|\lambda_{k,u}|^2\delta(\Delta_{jk}- \omega_k),
\end{equation}
where $\lambda_{k,u}$ accounts for the $u$th thermal bath coupled to a single-mode boson field with the frequency $\omega_k$, and energy gap
\begin{equation}
\Delta_{jk}=E_{j}-E_{k}.
\end{equation}

\noindent (ii) And the transition coefficients:
\begin{eqnarray}
    {S}^{jk}_q &=& {\langle}\phi_j|(\hat{\sigma}_++\hat{\sigma}_-)|\phi_k{\rangle},\\
    {S}^{jk}_c &=& {\langle}\phi_j|(\hat{a}^{\dag}+\hat{a})|\phi_k{\rangle}.
\end{eqnarray}
To satisfy the necessary conditions for the validity of the above master equation, we consider the Ohmic case:
\begin{equation}
    \gamma_u(\Delta_{jk})=\pi\alpha\Delta_{jk} e^{-|\Delta_{jk}|/\omega_{c}},
\end{equation}
where $\alpha$ is the coupling strength between the system and the environment and $\omega_{c}$ is the cutoff frequency of thermal baths.

Finally, the temperature of the bath is encoded in the Bose-Einstein distribution ($k_B = 1$)
\begin{equation}
n_{u}(\Delta_{jk}, T_{u})=\frac{1}{e^{\Delta_{jk}/T_{u}}-1}.
\end{equation}
It is important to emphasize that, when both reservoirs have the same temperature $T_a = T_{\sigma^-}$ (or only one of the subsystems is coupled to the reservoir) the system reaches a thermal equilibrium. Consequently, the dressed master equation steady-state solution of Eq.~\eqref{eq:dressed-me} results in the density matrix of the canonical ensemble [a straightforward numerical simulation proves that the Gibbs state is indeed the steady-state solution of Eq.~\eqref{eq:dressed-me}]
\begin{eqnarray}
	\hat{\rho}_\mathrm{ss}=\sum_k\frac{e^{-E_k/T}}{\mathcal{Z}}|\phi_k{\rangle}{\langle}\phi_k|,\label{eq:ss}
\end{eqnarray}
where $\mathcal{Z}=\sum_ke^{-E_k/T}$ is the partition function and the steady-state population is
\begin{eqnarray}
	P_k=\frac{e^{-E_k/T}}{\mathcal{Z}}.
\end{eqnarray}
Notice that, for unequal reservoir temperatures $T_a\neq T_{\sigma^-}$, the steady solution of the dressed master equation cannot be written in terms of the canonical ensemble as in Eq.~\eqref{eq:ss}. In what follows, we assume equal temperatures reservoir such that we can always extract the statistical properties of the quantum state at thermal equilibrium from Eq.~\eqref{eq:ss}.

\section{Quantum Correlation and Nonclassicality measures}\label{quantifiers}
The concept of quantumness (nonclassicality) is related to the impossibility of describing physical phenomena by a deterministic or probabilistic classical theory. To understand the impact of the system's light-matter anisotropy on the thermal quantum correlations and thermal nonclassicality, i.e. quantum correlations and nonclassicality that are present at thermal equilibrium, we thoroughly study the following quantities: (i) the zero-delay second-order correlation function $g^{(2)}(0)$, (ii) the bosonic field quadrature squeezing $\zeta^{2}$, (iii) a phase-space interference macroscopicity measure $\mathcal{I}(\rho)$ for the quantum state $\rho$, (iv) the negativity $\mathcal{N}(\rho)$, and (v) the quantum discord $\mathcal{D}(\rho)$. In the subsequent sections, we briefly explain the above quantities.

\subsection{Zero-delay second-order correlation function}
While the Poissonian and the super-Poissonian photon statistics of a light beam can be entirely explained in terms of a classical theory of light, the occurrence of sub-Poissonian photon statistics characterizes the quantumness of photonic states without a classical counterpart. It is also relevant to point out that, although this property does not consistently manifest in all quantum states of a field mode, a state can be classified as nonclassical when it is present. In this context, sub-Poissonian photon statistics serve as an authentic signature of the quantum nature of light.

Alternatively, one can classify the quantumness of a light beam by the study of the probabilities in measuring photons at a detector in a defined time interval $t_2 - t_1 \equiv \tau$, the so-called second-order correlation function $g^{(2)}(\tau)$. This definition classifies the light beam in a threefold fashion, namely: (i) $g^{(2)}(0)>1$ bunched light where photons populate themselves together (classically a chaotic description), (ii) $g^{(2)}(0)=1$ random photon stream (classically a coherent description), and (iii) $g^{(2)}(\tau)<1$ antibunched light where photons distribute separately (with no classical analogy)~\cite{carmichael1999statistical,carmichael2009statistical}, where we have considered an infinitesimal zero-delay time window $\tau \rightarrow 0$. Indeed, thermal photons emitted from non-interacting quantum modes or in the strong coupling regime is bunched ($g^{(2)}(0)=2$), being the standard example of an incoherent source of light
~\cite{carmichael1999statistical,carmichael2009statistical,glauber2006nobel}.

The conventional definition of the normalized zero-delay second-order correlation function is~\cite{glauber1963quantum}
\begin{equation}
	g^{(2)}(0) = \frac{\langle (\hat{a}^\dagger)^2(\hat{a})^2\rangle}{\langle \hat{a}^\dagger\hat{a} \rangle^2}.\label{eq:g2-aa}
\end{equation}
This quantity describes the probability of detecting two photons simultaneously ($\tau \rightarrow 0$), which is normalized by the probability of detecting two photons at once within a random photon source. Nonetheless, this definition holds for weak light-matter couplings, where the intracavity photons, described by $\hat{a}$, suffice to explain the observed photons correlation. On the other hand, in the USC regime, where the qubit system strongly dresses the bosonic field, the second-order correlation function $g^{(2)}(0)$ is derived from the input-output formalism as~\cite{rabl2011photon,ridolfo2012photon}
\begin{equation}
	G^{(2)}(0)=\frac{{\langle}(\hat{X}^-)^2(\hat{X}^+)^2{\rangle}}{{\langle}\hat{X}^-\hat{X}^+{\rangle}^2},~\label{eq:g2-x}
\end{equation}
where
\begin{eqnarray}~\label{xp}
	\hat{X}^+=-i\sum_{k>j}\Delta_{kj}X_{jk}|\phi_j{\rangle}{\langle}\phi_k|,
\end{eqnarray}
with $\hat{X}^-=(\hat{X}^+)^{\dag}$, $\Delta_{kj}=E_k-E_j$ being the energy gap, and $X_{jk}={\langle}\phi_j|(\hat{a}^\dag+\hat{a})|\phi_k{\rangle}$. Here, ${X}^+_{jk}$ describes the transition from the higher eigenstate $|\phi_k{\rangle}$ to the lower one $|\phi_j{\rangle}$. Notice that, in the weak light-matter interaction limit (i.e. $\lambda_i\ll 1$), the operator $\hat{X}^+$ is reduced to $\hat{X}^+=-i\omega\hat{a}$. Thus, the correlation function in Eq.~\eqref{eq:g2-x} simplifies to the conventional case.

One fundamental observation regarding the second-order correlation function in Eq.~\eqref{eq:g2-x} is that, in the eigenbasis of the AQRM Hamiltonian shown in Eq.~\eqref{eq:Hamiltonian-aqrm}, the dressed light-matter jump operator $\hat{X}$ provides the accurate expression for the average number of excitations in the ground state: $\langle\phi_0|\hat{X}^-\hat{X}^+|\phi_0\rangle = 0$. This is in contrast to the seemingly incorrect result $\langle\phi_0|\hat{a}^\dagger\hat{a}|\phi_0\rangle \neq 0$. Note that as the dressed ground state can be spanned in the bare light-matter basis $\{|g\rangle, |e\rangle, |n\rangle \}$, the dressed ground state will be composed of certain amount of virtual excitations ~\cite{garziano2013switching}. To convert the virtual excitations from the dressed picture into measurable photonic excitations, one can switch the interaction coupling strength on and off~\cite{garziano2013switching}.

Indeed, consider a quantum state in the dressed basis $|\phi_k\rangle$. Once the interaction is switched off, the state will still contain the excitations that can now be detected using the intracavity photon operator $\hat{a}$. Hence, $\langle \phi_k|\hat{a}^\dagger \hat{a}|\phi_k \rangle \neq 0$ will accurately describe the number of photonic excitations. It is worth noting that the non-adiabatic switch on/off needs to be of the order of the inverse of the relevant frequencies in the system, here ${\omega, \Delta}$~\cite{garziano2013switching}.

The conversion between virtual excitations and real photons via switching the interaction on and off is a consistent way to reconcile the correlation functions and other quantities that describe the thermal quantum correlations and nonclassicality in our work. Let us point out that the photons in the ground state of the USC models are virtual and cannot be detected \cite{de2014steady}, unless the coupling is suddenly switched off.

Throughout the paper, whenever a quantifier is defined using standard bosonic and spin operators, we refer to a scenario in which the coupling has been turned off after reaching thermal equilibrium.

\subsection{Quadrature squeezing}
The second-order correlation function fails as a nonclassical quantifier in the super-Poissonian regime. For instance, nonclassical Gaussian states, such as squeezed states may show either photon bunching or antibunching depending on whether the amplitude fluctuations are increased or reduced. To reveal the nonclassicality present in the steady state of the AQRM field mode, we compute the degree of squeezing. To do so, we introduce the generalized rotated field quadrature $\sqrt{2}\hat{x}_{\theta}=\hat{a}e^{-i\theta}+\hat{a}^{\dagger}e^{i\theta}$, where any two operators that differ by $\theta = \pi/2$ form a conjugate pair which satisfies the position ($\theta=0$) momentum ($\theta = \pi/2$) commutator relation. We consider the following squeezing parameter~\cite{lukvs1988principal}:
\begin{equation}
	\zeta^{2}=\underset{\theta\in(0,2\pi)}{\mathrm{min}}(\Delta\hat{x}_{\theta})^{2}.
\end{equation}
%which is the minimum value of $(\Delta X_{\theta})$ with respect
%to $\theta$, and 
Only a value of $\zeta^{2}<1$ indicates bosonic squeezing, whereas $\zeta^{2}=1$ corresponds to the photonic coherent state of radiation field. The minimization in the above definition can be easily obtained 
\begin{equation}
	\zeta^{2}=1+2\langle\hat{a}^{\dagger}\hat{a}\rangle-2|\langle\hat{a}^{2}\rangle|,
\end{equation}
where we have assumed that once the steady-state is reached, the light-matter interaction is switched off. Note that, Refs.~\cite{stassi2016output,noh2019output} studied the squeezing of output quadratures in the USC/DSC regime through the dressed light-matter jump operators for a vacuum input field, that is the thermal equilibrium case considered here. Interestingly, the authors found that any open system in its ground state cannot produce output squeezing, even if its ground state is a squeezed state. Since most of the squeezing in the AQRM is present in the ground state, these generalized measures leads to zero degree of squeezing in the thermal state if the coupling is on. For this reason we only focus in evaluating the degree of squeezing once the light-matter interaction is switched off, i.e., $\zeta^{2}$.

\subsection{Phase space interference measure}
To further study the nonclassicality of the AQRM steady-state, we consider a phase-space interference measure that quantifies the degree of macroscopicity of the state~\cite{lee2011quantification}, defined as:
\begin{equation}
	\mathcal{I} = - \frac{\pi}{2} \int \mathrm{d}p \mathrm{d}q W(q,p) \left( \frac{\partial^2}{\partial q^2} + \frac{\partial^2}{\partial p^2} + 1 \right),
\end{equation}
where $W(q,p)$ is the Wigner function~\cite{scully1997quantum}.
% that can be obtained from the following transformation of the density operator :
% \begin{equation}
% 	W(\alpha)=\frac{2}{\pi}{\rm Tr}[\hat{D}^{\dagger}(\alpha)\rho_{{\rm field}}\hat{D}(\alpha)(-1)^{\hat{a}^{\dagger}\hat{a}}]
% \end{equation}
% where $\hat{D}(\alpha)={\rm exp}(\alpha\hat{a}^{\dagger}-\alpha^{*}\hat{a})$ is the displacement operator and $\alpha=\alpha_{r}+i\alpha_{m}$ is a complex parameter.
The macroscopicity $\mathcal{I}$ takes the value of $0$ for classical states, while for pure quantum states such as superpositions of coherent states, NOON states, and Fock states, it corresponds to the average number of photons $\langle n \rangle$. In the case of the mixed state obtained by Eq.~\eqref{eq:ss}, the quantifier is constrained by the inequality $\mathcal{I}(\hat{\rho}_\mathrm{ss}) < \mathrm{Tr}(\hat{\rho}_\mathrm{ss}\hat{a}^{\dagger}\hat{a})$.

\subsection{Negativity}
To explore the quantum correlations between the light-matter constituents, we commence by calculating the quantum entanglement within bipartite systems. Among the various entanglement quantifiers available, we opt to use the negativity $\mathcal{N}(\rho)$~\cite{eisert1999comparison,zyczkowski1998volume,plenio2005logarithmic,lee2000partial,vidal2002computable}
\begin{equation}
	\mathcal{N}(\rho)=\frac{\Vert\rho^{T_{A}}\Vert_{1}-1}{2},
\end{equation}
where $\rho^{T_{A}}$ is the partial transpose of the quantum state $\rho$ with respect to subsystem A, $\Vert Y\Vert_{1}={\rm Tr}|Y|={\rm Tr}\sqrt{Y^{\dagger}Y}$ is the trace norm or the sum of the singular value the operator $Y$. Equivalently, the negativity can be computed as $\mathcal{N}(\rho)=1/2\sum_i(|\varepsilon_i|-\varepsilon_i)$, where $\varepsilon_i$ are the eigenvalues of the partially transposed light-matter density matrix $\rho$. Note that $\mathcal{N}(\rho)=0$ corresponds to separable (not entangled) quantum states.

\subsection{Quantum discord}
Quantum entanglement is not the sole manifestation of quantum correlations. In fact, quantum discord (QD) represents another measure of potential quantum correlations within a quantum system. This measure of correlation emerges from the observation that two classically equivalent methods of defining mutual information yield different outcomes within the quantum domain~\cite{ollivier2001quantum}. The QD of two subsystems $\mathcal{A}$ and $\mathcal{B}$ can be expressed as~\cite{ollivier2001quantum}:
\begin{equation}
\mathcal{D}(\hat{\rho}_{\mathcal{AB}})=S(\hat{\rho}_{A})-S(\hat{\rho}_{\mathcal{AB}})+{{\rm min}}_{\{\Pi_{j}^{\mathcal{A}}\}}S(\hat{\rho}_{\mathcal{B}|\{\Pi_{j}^{\mathcal{A}}\}}),
\end{equation}
% \begin{equation}
% 	\begin{aligned}
% 		\mathcal{D}(\hat{\rho}_{\mathcal{AB}})& :=\mathcal{I}(\hat{\rho}_{\mathcal{AB}})-{{\rm max}}_{\{\Pi_{j}^{\mathcal{A}}\}}\mathcal{J}_{\{\Pi_{j}^{\mathcal{A}}\}}(\hat{\rho}_{\mathcal{AB}}),\\
% 		& =S(\hat{\rho}_{A})-S(\hat{\rho}_{\mathcal{AB}})+{{\rm min}}_{\{\Pi_{j}^{\mathcal{A}}\}}S(\hat{\rho}_{\mathcal{B}|\{\Pi_{j}^{\mathcal{A}}\}}),
% 	\end{aligned}
% \end{equation}
where $S(\hat{\rho}_{i})=-{\rm Tr}\hat{\rho}_{i}{\rm log}\hat{\rho}_{i}$ is the von Neumann entropy for the reduced density matrix, and 
% $\mathcal{J}_{\{\Pi_{j}^{A}\}}(\hat{\rho})$ is the quantum mutual information:
% \begin{equation}
% 	\mathcal{J}_{\{\Pi_{j}^{A}\}}(\hat{\rho}_{\mathcal{AB}})=S(\hat{\rho}_{\mathcal{B}})-S(\hat{\rho}_{\mathcal{B}|\{\Pi_{j}^{\mathcal{A}}\}})
% \end{equation}
$S(\hat{\rho}_{\mathcal{B}|\{\Pi_{j}^{A}\}}$ is the entropy conditioned through performing measurements on the $\mathcal{A}$ system, defined as:
\begin{equation}
S(\hat{\rho}_{\mathcal{B}|\{\Pi_{j}^{A}\}})=\sum_{j}p_{j}S((\{\Pi_{j}^{\mathcal{A}}\}\otimes I^{\mathcal{B}})\hat{\rho}_{\mathcal{AB}}(\{\Pi_{j}^{\dagger \mathcal{A}}\}\otimes I^{\mathcal{B}})/p_{j}).
\end{equation}
In the above, $\Pi_{j}^{\mathcal{A}}$ is a von Neumann projection operator on subsystems $\mathcal{A}$ and $p_{j}={\rm Tr}(\{\Pi_{j}^{\mathcal{A}}\}\otimes I^{\mathcal{B}}\hat{\rho}_{\mathcal{AB}})$ is the probability with measurement outcome $j$. Note that, one of the most distinct differences between QD from entanglement is that QD can be non-zero for certain separable states~\cite{PhysRevA.77.042303, Coto_2017, Ali_2010}.

\section{Quantumness at thermal equilibrium}
\label{results}

%\subsection{Effect of light-matter coupling strength}

\begin{figure}
\centering \includegraphics[width=\linewidth]{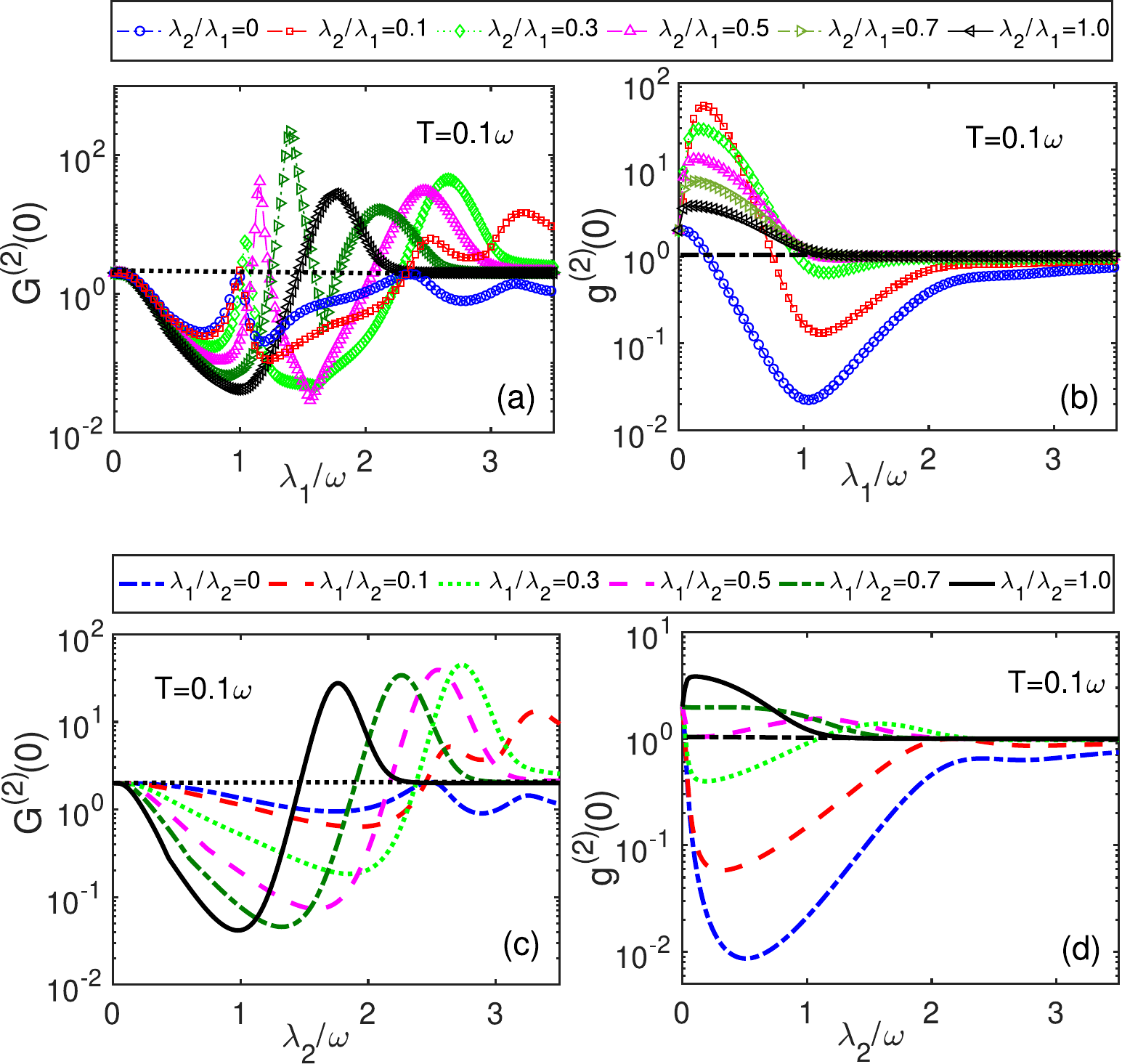}
\caption{
Second-order correlation functions: Panels (a) and (c) show $G^{(2)}(0)$ as a function of the coupling strength $\lambda_{1}$ ($\lambda_{2}$) for several choices of coupling ratios $\lambda_{2}/\lambda_{1}$ ($\lambda_{1}/\lambda_{2}$). The black dotted horizontal line indicates $G^{(2)}(0)=2$. Panels (b) and (d) depict $g^{(2)}(0)$ as a function of $\lambda_{1}$ ($\lambda_{2}$) for different coupling ratios $\lambda_{2}/\lambda_{1}$ ($\lambda_{1}/\lambda_{2}$). The black dashed–dotted horizontal line corresponds to $g^{(2)}(0)=1$. The other system parameters are the same as in Fig.~\ref{fig1}.
}~\label{fig2}
\end{figure}

We commence by examining the impact of the light-matter coupling strengths, $\lambda_{1}$ and $\lambda_{2}$, on the generation of long-lived quantum correlations in the AQRM. In Figs.~\ref{fig2}(a)-(c), we plot the second-order correlation function using the dressed jump operator $G^{(2)}(0)$ as a function of the light-matter coupling strength $\lambda_i$ ($i=1, 2$) for several coupling ratios $\lambda_i/\lambda_j$ ($i\neq j$) and a given temperature $T=0.1\omega$. As the panels (a) and (c) show, $G^{(2)}(0)$ generally transitions from photon antibunching ($G^{(2)}(0)<1$) to bunching ($G^{(2)}(0)>1$) as the light-matter coupling strength increases. In the limit of the DSC regime, i.e. $\lambda \gg \{\omega,\Delta\}$, the correlation function takes the value $G^{(2)}(0)=2$ which corresponds to thermal photon emission (regardless of the coupling ratios). In Figs.~\ref{fig2}(b)-(d), we depict the second-order correlation function using the bare jump operator $g^{(2)}(0)$ as a function of $\lambda_i$ ($i=1, 2$) for several coupling ratios $\lambda_i/\lambda_j$ ($i\neq j$) and the same temperature $T=0.1\omega$. As the panels (b) and (d) show, the second-order correlation function $g^{(2)}(0)$ exhibits nonclassical features only for a strong degree of anisotropy (when the coupling strength ratio is small), i.e $\lambda_{2}/\lambda_{1}\leqslant 0.3$ and $\lambda_{1}/\lambda_{2}\leqslant 0.3$, respectively. Interestingly, as the coupling ratios $\lambda_i/\lambda_j$ increase, the nonclassical features (photonic antibunching effect) are completely suppressed ($g^{(2)}(0)>1$). Furthermore, with an increase in the light-matter coupling, $g^{(2)}(0)$ approaches unity regardless of the choice of coupling ratios. In other words, the thermal emission now displays the same statistical behavior as a coherent state. Note that in computing $g^{(2)}(0)$ using the bare bosonic field, we make the assumption that the coupling strength has been switched off~\cite{garziano2013switching} once the system has reached its thermal equilibrium. Notably, there is a significant distinction between the scenario in which the correlation function is computed from $G^{(2)}(0)$ and the one obtained when the light-matter coupling is switched off, resulting in $g^{(2)}(0)$. This subtlety demonstrates the intricate impact of virtual photonic excitations on the thermal state properties of the AQRM.

%==========================================

We can gain a clearer understanding of the second-order correlation function $G^{(2)}(0)$ by considering the energy level crossing and its associated change in parity, as similarly discussed in Ref.~\cite{xu2020two}. 

\begin{figure}
\centering \includegraphics[width=\linewidth]{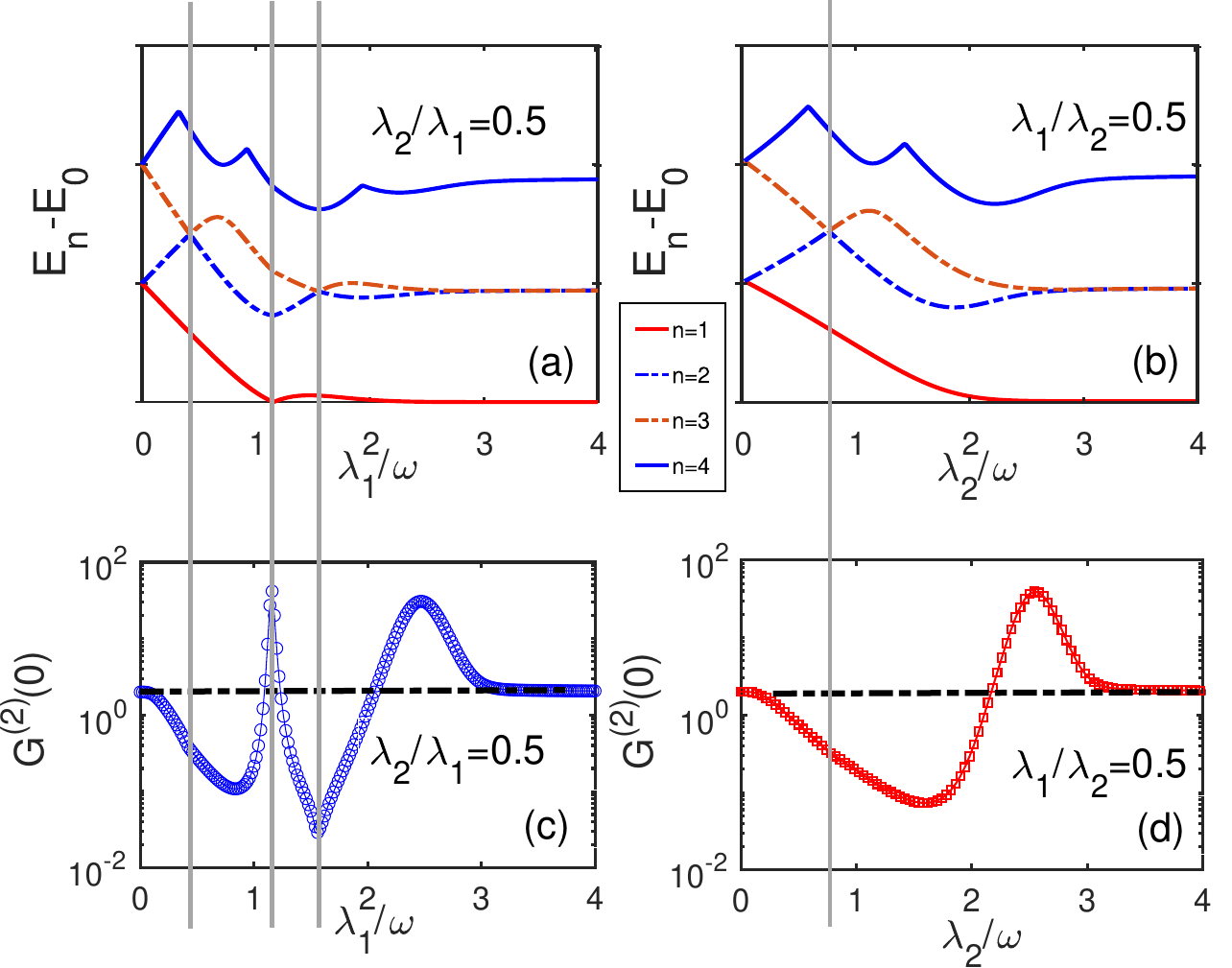}
\caption{Relationship between energy level crossing and the behavior of the second-order correlation function $G^{(2)}(0)$. In panels (a) and (b), we redraw the energy gap $E_n-E_0$ of the AQRM as a function of $\lambda_1$ ($\lambda_2$) for $\lambda_2/\lambda_1=0.5$ ($\lambda_1/\lambda_2=0.5$), see more cases of coupling ratios in Fig.~\ref{fig1}. In panels (c) and (d), we show the particular case of $G^{(2)}(0)$ as a function of $\lambda_1$ ($\lambda_2$) for $\lambda_2/\lambda_1=0.5$ ($\lambda_1/\lambda_2=0.5$). The vertical gray lines evidence the correspondence between sharp peaks of $G^{(2)}(0)$ and the closing of energy gaps. Other system parameters are chosen as in Fig.~\ref{fig1}.
}~\label{fig3}
\end{figure}

Consider the particular case of $G^{(2)}(0)$ as a function of $\lambda_1$ for a fixed coupling ratio $\lambda_2/\lambda_1 = 0.5$, as redrawn in Fig.~\ref{fig3}(c) for clarity. The non-analytical (sharp) peaks result from the energy level crossing depicted in Fig.~\ref{fig3}(a) (see Fig.~\ref{fig1} for other choices of parameters). Indeed, the closing of the energy gap between higher excited states, namely the parity change between the second ($E_2$) and the third ($E_3$) energy levels are responsible for the anti-bunching peaks shown in Fig.~\ref{fig3}(c). Furthermore, the closing of the energy gap between the ground state ($E_0$) and its first energy level ($E_1$) gives rise to the pronounced photonic bunched peak in Fig.~\ref{fig3}(c) (around the value of $\lambda_1\sim\omega$). Remarkably, the closing of the energy gap between $E_1$ and its ground state $E_0$ reopens as the coupling strength increases. This cycle of opening, closing, and reopening dynamics is responsible for the transition between antibunched-bunched-antibunched behavior of $G^{(2)}(0)$ shown in Fig.~\ref{fig3}(c). A similar analysis can be conducted for Fig.~\ref{fig3}(d), in which we present $G^{(2)}(0)$ as a function of $\lambda_2$ for a fixed coupling ratio $\lambda_1/\lambda_2 = 0.5$. In contrast to the previous scenario, no non-analytic behavior is observed. This is because although a crossing of energy levels between the second and third eigenenergies occurs, the energy gap between the ground state and the first excited state monotonically vanishes as the coupling strength increases. As a result, a smooth transition towards a photonic bunching effect takes place.

%%==========================================

%In the regime $\lambda_{1}\in(0.42\omega,1.15\omega)$, there is an crossing between energy levels $E_{2}$ and $E_{3}$ , the corresponding parity is exchanged (see the dashed blue line and dashed red line in Figure 1(c)). The two-photon correlation function is changed to $G^{(2)}(0)\approx P_{2}(\Delta_{21}X_{21})^{2}/[P_{1}^{2}(\Delta_{10}X_{10})^{2}]$

%In the regime $\lambda_{1}\in(1.15\omega,1.55\omega)$, the energy levels $E_{1}$ and $E_{0}$ begin to overlap but are not completely degenerate. (see the blue solid line  and red dashed line  in figure 1(c)). The effective two-photon correlation function is given by  $G^{(2)}(0)\approx [P_{3}(\Delta_{31}X_{31}\Delta_{10}X_{10})^{2}+P_{3}(\Delta_{32}X_{32}\Delta_{20}X_{20})^{2}]/[P_{1}^{2}(\Delta_{10}X_{10})^{4}]$, by further increasing the coupling strength, we can also get the approximate expression of  $G^{(2)}(0)$ according to the above analysis method. For the $\lambda_{1}/\lambda_{2}=0.5$ as seen in Fig.~\ref{fig3}(b), (d)(f) and other coupling ratios, the approximate expression of $G^{(2)}(0)$ can also be obtained from the perspective of parity change.

\begin{figure}
\centering \includegraphics[width=\linewidth]{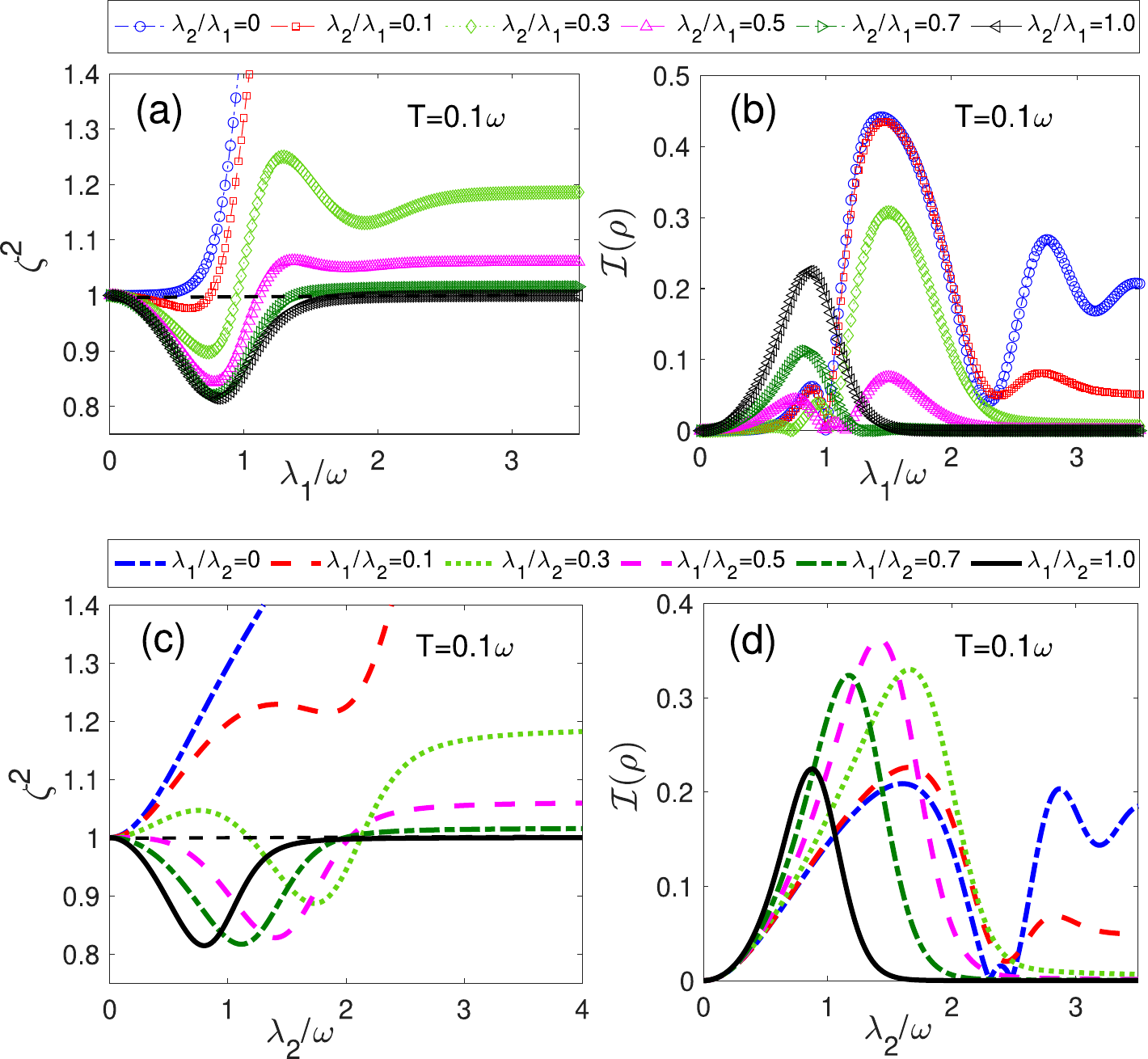}
\caption{Panel (a) [(c)] shows the squeezing parameter $\zeta^2$ as a function of $\lambda_1$ [$\lambda_2$] for different coupling ratios $\lambda_2/\lambda_1$ [$\lambda_1/\lambda_2$]. The black dashed horizontal line corresponds to $\zeta^{2}=1$. Panel (b) [(d)] depicts the macroscopicity interference-based measure $\mathcal{I}(\rho)$ as a function of $\lambda_1$ [$\lambda_2$] for several coupling ratios $\lambda_2/\lambda_1$ [$\lambda_1/\lambda_2$]. We consider a given temperature $T=0.1\omega$ and other system parameters are the same as in Fig.~\ref{fig1}.
}~\label{fig4}
\end{figure}

To investigate the degree of photonic squeezing, in Fig.~\ref{fig4}(a) [Fig.~\ref{fig4}(c)], we plot the squeezing parameter $\zeta^2$ as a function of $\lambda_1$ [$\lambda_2$] for several coupling ratios $\lambda_{2}/\lambda_{1}$ [$\lambda_{1}/\lambda_{2}$] for a given temperature $T=0.1\omega$. As the figures show, there is a clear degree of squeezing $\zeta^2<1$ of the AQRM steady-state (after the interaction has been switched off) for certain coupling ratios and a specific window of light-matter coupling. Note that for the JCM ($\lambda_{2}/\lambda_{1}=0$), no squeezing is achieved, i.e. $\zeta^2 \geq 1$, for all values of $\lambda_1$. Conversely, for the QRM ($\lambda_{2}=\lambda_{1}$), the degree of squeezing tends to a photonic coherent field as the light-matter coupling strength goes towards the USC/DSC regime. Notably, no photonic squeezing occurs for the high anisotropy regime ($\lambda_i/\lambda_j \ll 1$, $i \neq j$). For instance, consider $\lambda_{2}/\lambda_{1}=0.1$ and $\lambda_{1}/\lambda_{2}=0.1$ in Figs.~\ref{fig4}(a) and \ref{fig4}(c), respectively. As seen from the figures, no relevant photonic squeezing is achieved with the increasing of the light-matter coupling. This nonclassicality measure is in stark contrast with the second-order correlation function $g^{(2)}(0)$ (we recall that both cases are evaluated by turning the coupling off) for which it presents high degree of nonclassicality in the region of high anisotropy.

%Furthermore, as a complementary measure, we study the photon squeezing as a measure of nonclassicality. We plot the quadrature squeezing $\zeta^{2}$ in steady state as a function of the coupling strength $\lambda_{1}$ ($\lambda_{2}$) with different coupling ratios $\lambda_{2}/\lambda_{1}$($\lambda_{1}/\lambda_{2}$) in Fig.~\ref{fig4}(a)( Fig.~\ref{fig4}(c)). There is no photon compression($\zeta^{2}\geq1$ ) in any coupling regime for the JC model ($\lambda_{2}/\lambda_{1}=0$). However, to a relatively strong coupling regime of $\lambda_{1}$, the increase of coupling $\lambda_{2}/\lambda_{1}>0$ helps to improve the Squeezing effect of photon ($\zeta^{2}<1$ ) and broaden the coupling regime where the squeeze effect of the photon appears. With further increase in the coupling strength $\lambda_{1}$ ($\lambda_{2}$) in Fig.~\ref{fig4}(a) (Fig.~\ref{fig4}(c)) entering the deep strong regime, the photon compression quickly disappears, regardless of the coupling rate.

%==========================================
%==========================================

To further evidence quantum features of the steady-state AQRM field that may have not been captured by previous quantifiers, in Fig.~\ref{fig4}(b) [Fig.~\ref{fig4}(d)], we show the macroscopicity interference-based measure $\mathcal{I}(\rho)$ as a function of $\lambda_{1}$ [$\lambda_{2}$] for different coupling ratios $\lambda_{2}/\lambda_{1}$ [$\lambda_{1}/\lambda_{2}$] and $T=0.1\omega$. As the figures show, the higher degree of interference-based measure generally occurs in the coupling region $1 < \lambda_i/\omega < 2$ ($i=1,2$) for most of the coupling ratios of anisotropies, where quantum squeezing did not capture such a nonclassical behavior, yet it appears to be closer to capturing the second-order correlation quantifier. Interestingly, as shown in Fig.~\ref{fig4}(b) $\mathcal{I}(\rho)$ increases with increasing anisotropy, thus exhibiting clear quantum signatures in the DSC regime. In contrast, in Fig.~\ref{fig4}(d) $\mathcal{I}(\rho)$ is maximum for $\lambda_{1}/\lambda_{2}=0.5$, which is a maximum of nonclassicality that is hidden in the previous measures. 

%To deepen the study of the quantum characteristics of photons that are robust to temperature and that may not be captured by the previous quantifiers, we investigate the macroscopicity $\mathcal{I}(\rho)$ as a function of the coupling strength $\lambda_{1} $ ($\lambda_{ 2}$) with different coupling rates $\lambda_{2}/\lambda_{1}$($\lambda_{1}/\lambda_{2}$) in Fig.~\ref{fig4} (b)( Fig.~\ref{fig4}(d)).

\begin{figure}
\centering \includegraphics[width=\linewidth]{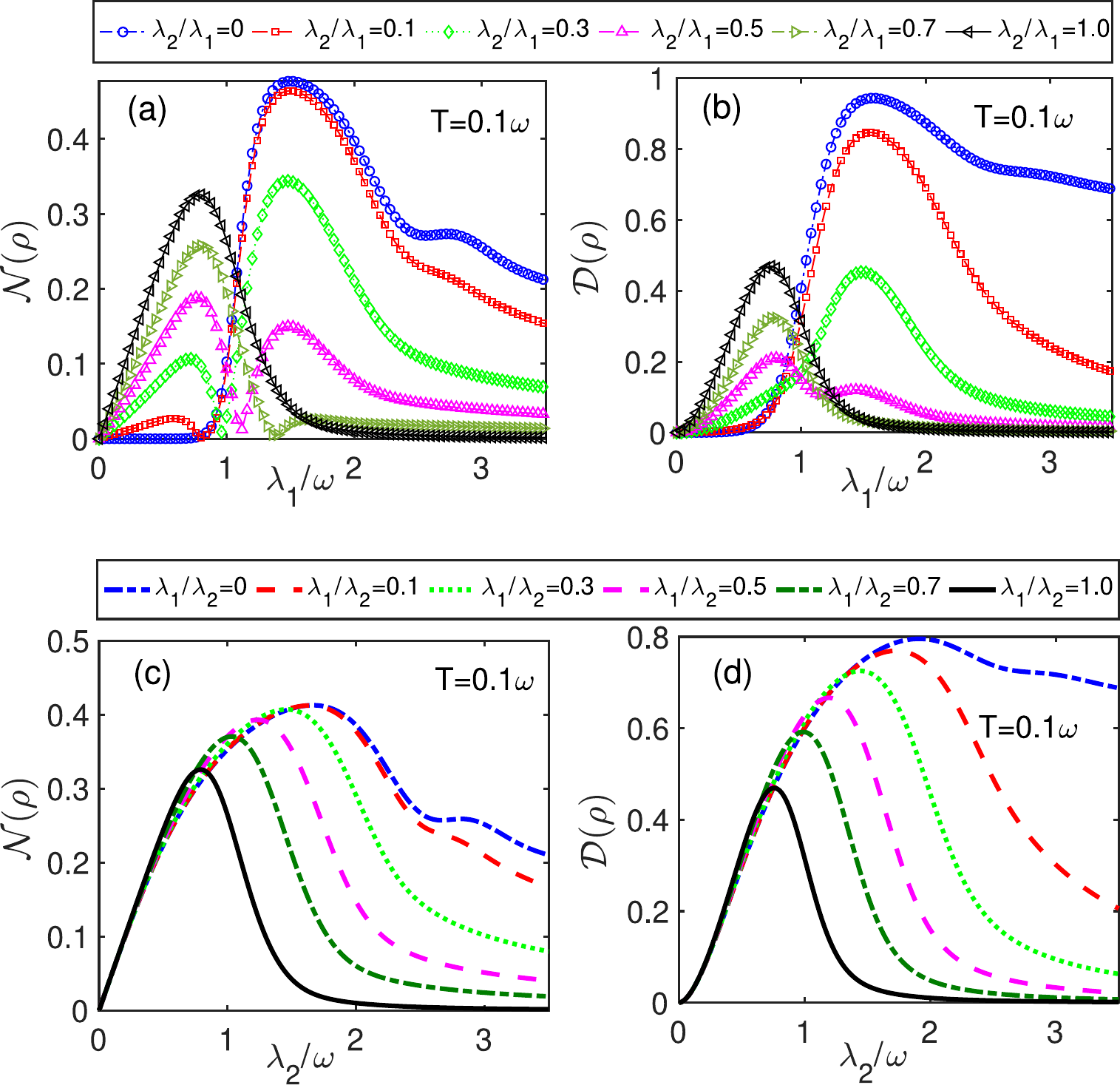}
\caption{Panel (a) [(c)] shows the steady-state negativity $\mathcal{N}(\rho)$ as a function of $\lambda_1$ [$\lambda_2$] for different coupling ratios $\lambda_2/\lambda_1$ [$\lambda_1/\lambda_2$]. Panel (b) [(d)] depicts the steady-state quantum discord $\mathcal{D}(\rho)$ as a function of $\lambda_1$ [$\lambda_2$] for several coupling ratios $\lambda_2/\lambda_1$ [$\lambda_1/\lambda_2$]. We consider a given temperature $T=0.1\omega$ and other system parameters are the same as in Fig.~\ref{fig1}.}~\label{fig5}
\end{figure}

Now, we study the thermal quantum correlations between the photons and the qubit through the quantifiers of negativity $\mathcal{N}(\rho)$ and quantum discord $\mathcal{D}(\rho)$. In Fig.~\ref{fig5}(a), we plot the steady-state negativity $\mathcal{N}(\rho)$ as a function of $\lambda_1$ for several coupling ratios $\lambda_2/\lambda_1$ for $T=0.1\omega$. As the figure shows, at some coupling ratios such as $\lambda_{2}/\lambda_{1}=0.1, 0.3, 0.5, 0.7$, the negativity shows a clear dip near to zero between two maxima. This minima near zero corresponds to the energy level crossing of the ground-state and the first excited state, see Figs.~\ref{fig1}(b)-(d). In the case of the JCM ($\lambda_{2}/\lambda_{1}=0$), the steady-state negativity $\mathcal{N}(\rho)$ is nearly zero for coupling strengths $\lambda_{1}\leq 0.7\omega$. However, as the coupling strength $\lambda_{1}$ increases, the negativity gradually rises until it reaches the highest degree of entanglement compared to any other coupling rate. In the case of the QRM ($\lambda_{2}/\lambda_{1}=1$), $\mathcal{N}(\rho)$ exhibits a single maximum with the highest degree of entanglement for coupling strength $\lambda_1 < \omega$. In Fig.~\ref{fig5}(c), we depict the steady-state negativity $\mathcal{N}(\rho)$ as a function of $\lambda_2$ for different coupling ratios $\lambda_1/\lambda_2$ at a fixed temperature of $T=0.1\omega$. As observed in the figure, the $\mathcal{N}(\rho)$ exhibits a smooth behavior without a sudden drop near zero. This is because, in contrast to the previous scenario, there is no energy level crossing between the ground state and the first excited state, as shown in Figs.~\ref{fig1}(g)-(i). While the steady-state negativity of Fig.~\ref{fig5}(c) is qualitatively similar to the steady-state quantum discord $\mathcal{D}(\rho)$ of Fig.~\ref{fig5}(d), the situation differs for Figs.~\ref{fig5}(a)-(b). Indeed, in Fig.~\ref{fig5}(b), we plot the quantum discord $\mathcal{D}(\rho)$ as a function of $\lambda_1$ for several coupling ratios $\lambda_2/\lambda_1$ for $T=0.1\omega$. As the figure shows, its overall profile is similar to that of Fig.~\ref{fig5}(a), this means that almost all non-local quantum correlations are due to entanglement. However, unlike negativity shown in Fig.~\ref{fig5}(a), at the crossing of the ground state and the first excited state the quantum discord $\mathcal{D}(\rho)$ does not decrease near to zero for some coupling strengths (e.g. $\lambda_{2}/\lambda_{1}=0.1, 0.3, 0.5, 0.7$). Thus, we can infer: (i) all non-local quantum correlations are in the form of entanglement; (ii) the quantum correlations increase with anisotropy, and (iii) for high anisotropy they increase with the coupling strength presenting its maximum in the DSC regime.

It is worth noting that the system's behavior for increasing light-matter coupling, e.g., $\lambda_i \gtrsim 3\omega$ ($i=1,2$), qualitatively resembles the system's behavior for the decoupled light-matter case, i.e., $\lambda_i = 0$. Such a qualitative correspondence implies a trivial behavior of the quantifiers for very large light-matter interaction strengths, $\lambda_i \gtrsim 3\omega$ ($i=1,2$), and for anisotropies with values exceeding $\lambda_i/\lambda_j \geq 0.3$, where $i \neq j$, see Fig.~\ref{fig1}. This behavior can be understood as, in the depths of the DSC regime, the spectrum becomes harmonic again (see Fig.~\ref{fig1} for $\lambda_i > 3\omega$), with the dressed states forming product states between displaced Fock photonic states and atomic states that are $x$-polarized, leading to $G^{(2)}(0) = 2$. The nearly equal energy gaps (quasi-harmonic spectrum) and the eigenstates being product states in the DSC regime suggest that: i) no degree of squeezing and macroscopicity, ii)  the vanishing of quantum entanglement and quantum discord at all temperatures should be expected (as shown in Figs.~\ref{fig2} to~\ref{fig5}). 

%==========================================
%==========================================

\section{robustness of thermal quantumness at higher temperatures}
%Effect of finite temperatures of thermal baths

As temperatures rise, thermal fluctuations increase, thereby modifying the AQRM steady-state as described in Eq.~\eqref{eq:ss}. On one hand, higher temperatures result in an increased expansion of states in the polariton basis, which may potentially lead to more frequent energy level crossings among higher AQRM eigenstates. On the other hand, these thermal fluctuations may cause the system to lose its coherence, potentially having a detrimental impact on the AQRM steady-state. To explore this non-trivial scenario, which involves a trade-off between increased energy level interplay and the influence of thermal fluctuations, we investigate how higher finite thermal bath temperatures affect the generation of nonclassical and quantum correlations in the AQRM, while considering all the above quantifiers.

%%==========================================
\begin{figure*}
\centering \includegraphics[width=0.7\linewidth]{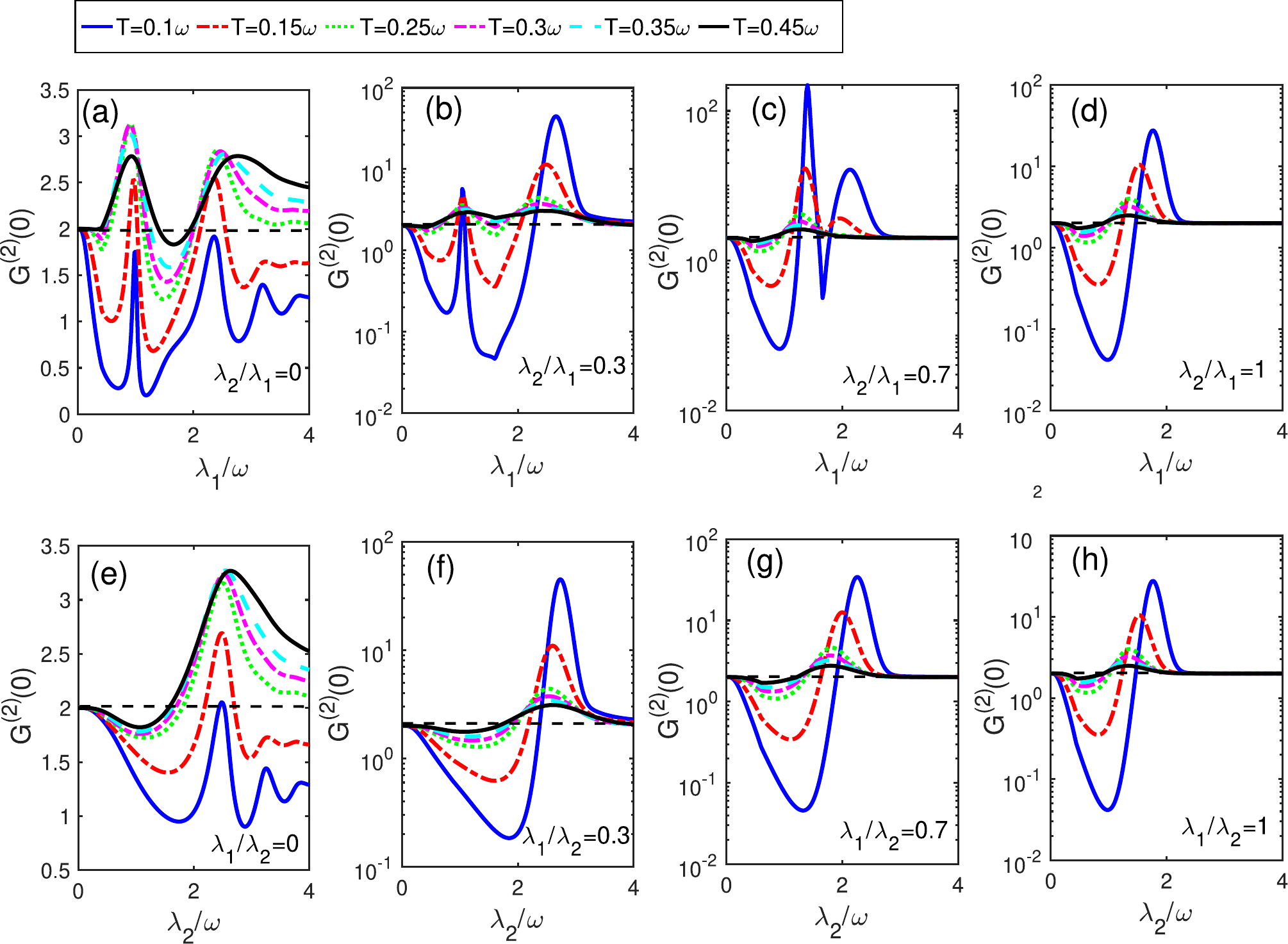}
\caption{Second-order correlation function $G^{(2)}(0)$ as a function of the light-matter coupling $\lambda_i$ ($i = 1,2$) for several choices of coupling ratios $\lambda_i/\lambda_j$ ($i\neq j$) and different temperatures $T$. The black dashed horizontal line corresponds to $G^{(2)}(0)=2$. Other system parameters are as in Fig.~\ref{fig1}.
}~\label{fig6}
\end{figure*}
%%==========================================

In Fig.~\ref{fig6}, we illustrate the second-order correlation function $G^{(2)}(0)$ as a function of the light-matter coupling $\lambda_i$ ($i = 1,2$) for several choices of coupling ratios $\lambda_i/\lambda_j$ ($i\neq j$) and different temperatures $T$. As evident from the figures, the higher the anisotropy, the more robust the quantum correlations are against temperature effects. Moreover, it exhibits a larger quantum region as a function of the coupling strengths. Notably, the correlation function $G^{(2)}(0)$ is prone to thermal decoherence, as signatures of nonclassicality (antibunching) survives only at temperatures as high as $T\sim 0.15\omega$, as depicted in all panels of Fig.~\ref{fig6}. For higher temperatures, the antibunching effect is largely suppressed with increasing temperature for the same values of the coupling ratio, suggesting that thermodynamic fluctuations consistently destroy nonclassical features in high-temperature regions.

%==========================================
\begin{figure*}
\centering \includegraphics[width=\linewidth]{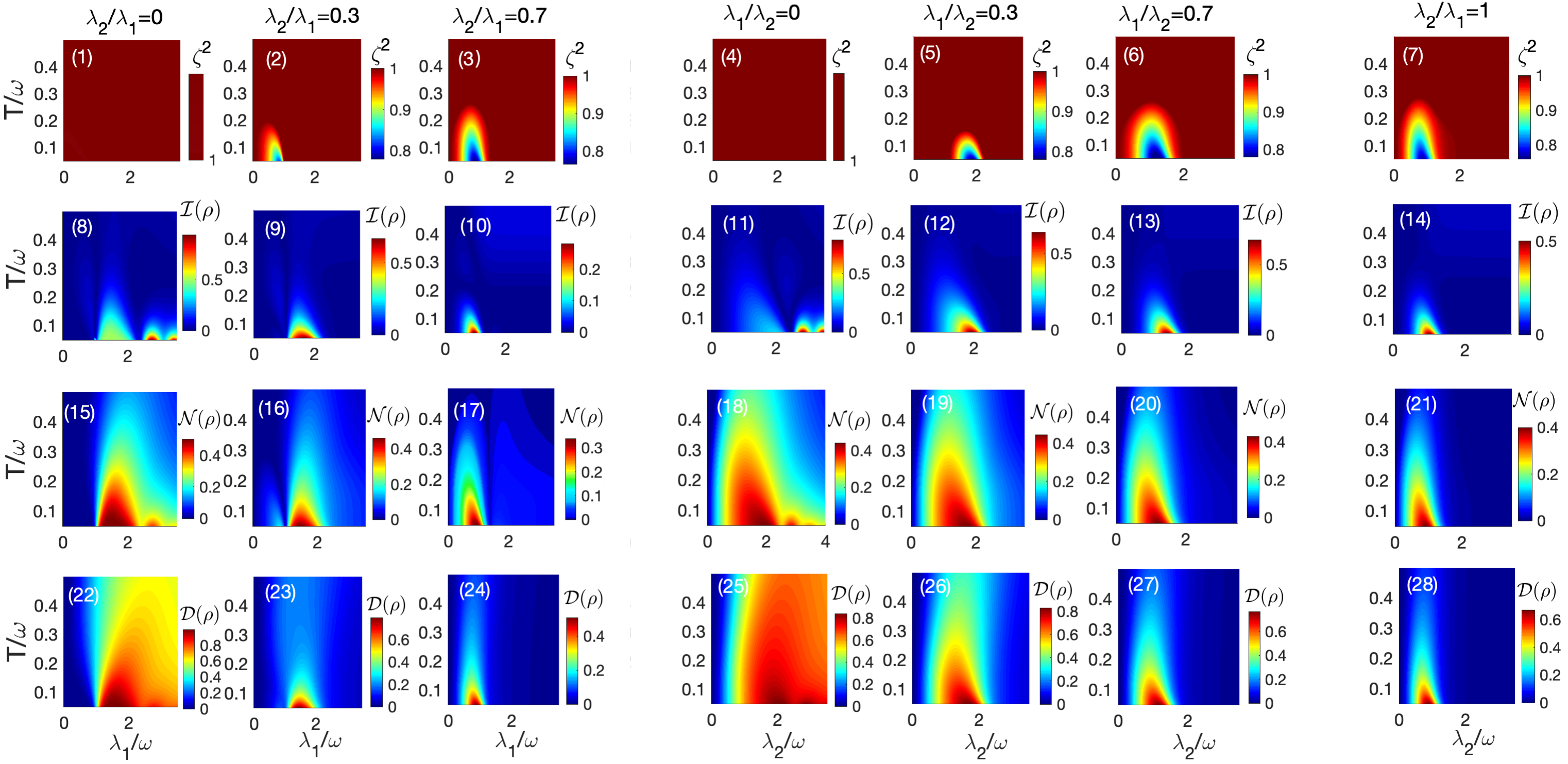}
\caption{The phase diagram of the squeezing parameter $\zeta^2$, macroscopicity $\mathcal{I}(\rho)$, negativity $\mathcal{N}(\rho)$, and quantum discord $\mathcal{D}(\rho)$, as functions of the light-matter coupling $\lambda_i$ ($i=1,2$) and the temperature $T$ for different coupling ratios $\lambda_i/\lambda_j$ ($i\neq j$). Other system parameters are as in Fig.~\ref{fig1}.
}~\label{fig7}
\end{figure*}
%==========================================

In Fig.~\ref{fig7}, we show the temperature effects on the four quantifiers, namely $\zeta^2$, $\mathcal{I}(\rho)$, $\mathcal{N}(\rho)$, and $\mathcal{D}(\rho)$, as functions of the light-matter coupling $\lambda_i$ ($i=1,2$) and the temperature $T$ for different coupling ratios $\lambda_i/\lambda_j$ ($i\neq j$). 

In the first row of Fig.~\ref{fig7}, we plot the squeezing parameter $\zeta^{2}$ as functions of $\lambda_i$ ($i=1,2$) and $T$ for different $\lambda_i/\lambda_j$ ($i\neq j$) values. Specifically, as observed in panels (1) and (4) of Fig.~\ref{fig7}, no squeezing effect is achieved in the specific cases of the JCM ($\lambda_2=0$) and the AJCM ($\lambda_1=0$). However, for other coupling ratios, significant squeezing effects are observed at temperatures around $T \approx 0.3\omega$. Such a high degree of squeezing is attained by decreasing the anisotropy, making the Rabi model the ideal scenario for obtaining this resource. 

In the second row of Fig.~\ref{fig7}, we depict the macroscopicity interference-based measure $\mathcal{I}(\rho)$ as functions of $\lambda_i$ ($i=1,2$) and $T$ for different $\lambda_i/\lambda_j$ ($i\neq j$) values. Unlike in the case of the squeezing parameter, the macroscopicity interference-based measure for the JCM and AJCM cases is non-zero, as shown in panels (8) and (11) of Fig.~\ref{fig7}. It's worth noting that macroscopicity takes non-zero values over a broader range of coupling strengths compared to other ratios at low temperatures $T < 0.1\omega$. However, as the figure shows, this quantifier is more susceptible to thermal fluctuations originating from the thermal reservoir and practically disappears for temperatures $T > 0.1\omega$.

Lastly, in the third and fourth rows of Fig.~\ref{fig7}, we depict steady-state quantum entanglement $\mathcal{N}(\rho)$ and quantum discord $\mathcal{D}(\rho)$. Both quantifiers exhibit qualitatively similar behaviors. Similar to the macroscopicity interference-based measure, both quantifiers are more pronounced over a broader range of coupling strengths and temperatures. As the ratio $\lambda_2/\lambda_1$ ($\lambda_1/\lambda_2$) increases, the range in which both quantifiers appear gradually narrows. In the case of $0 < \lambda_2/\lambda_1 < 1$, as $\lambda_1$ changes, quantum entanglement $\mathcal{N}(\rho)$ appears in two regions, with a value of zero between these two regions. This result is due to the energy level crossing between the ground state and the first excited state. However, for quantum discord $\mathcal{D}(\rho)$, only one region appears. Interestingly, in the case of high anisotropy, entanglement persists up to temperatures $T \approx 0.3\omega$, while quantum discord demonstrates greater robustness against thermal fluctuations up to temperatures $T \lesssim 0.4\omega$.

Our theoretical proposal to reach nonclassical equilibrium states is not just of academic interests and also meets the present-day nontrivial experimental challenges, mainly involving the coupling strength and low temperature environments. On the one hand, the physical model considered is general and has been implemented in a large number of systems, such as cavity quantum electrodynamics~\cite{grimsmo2013cavity}, and superconducting circuits~\cite{xie2014anisotropic}. Among them, the USC regime has been experimentally demonstrated for more than a decade~\cite{anappara2009signatures} ($\lambda/\omega > 0.1$). Since then, the USC regime has been achieved in several other systems~\cite{forn2019ultrastrong}. On the other hand, the DSC regime is within the reach of experimental techniques, notably in Ref.~\cite{yoshihara2017superconducting} in circuit quantum electrodynamics setup implementing the QRM, for the highest reported strength of light-matter interaction of $\lambda/\omega \approx 1.34$ with temperature of  approximately 45 mK ($k_{B}T/\hbar\omega \approx 0.16$), where thermal entanglement was also confirmed.

\section{Conclusion}
\label{conclusion}
In this work, we consider the steady-state of the AQRM  by solving a dressed master equation at thermal equilibrium. Such a master equation in the dressed basis is valid for any light-matter coupling strength, and where the resulting steady-state is indeed a Gibbs state. Our findings are two-fold: (i) we investigated the generation of quantum correlations and nonclassicality in the AQRM steady-state at thermal equilibrium. Through a comprehensive analysis of various measures of quantum correlation and nonclassicality, we demonstrate that quantum effects persist across a broad range of anisotropies and coupling strengths, even at moderate thermal equilibrium temperatures; and (ii) such quantum features significantly emerge in the USC regime and at the onset of the DSC regime, while practically vanishing in the deeper DSC regime. In the dressed picture, this results in virtual field excitations within the system. Notably, we show a significant difference in the second-order correlation function when computed using dressed jump operators compared to using bare field operators (i.e., when the light-matter coupling is assumed to be switched off after the system reaches thermal equilibrium). Both the difference between these quantifiers and the persisting quantum features of the AQRM steady-state at high temperatures, could allow for experimental approaches to study the impact of virtual excitations in the USC regime. In particular, as the quantum features emerging from the system avoid the need for demanding ground-state cooling.

%To summarize, we investigated the nonclassicality present in AQRM that survives thermalization and can be harvested as a quantum resource. We employ a quantum Markovian master equation to describe the evolution of AQRM that is valid for all coupling regimes and verified that the steady state is a Gibbs equilibrium state. In addition, through a set of complementary measures, we quantify the different quantum aspects. Our theoretical proposal to reach nonclassical equilibrium states presents nontrivial experimental challenges, mainly involving the coupling strength and low temperature environments. The bright side is that the physical model considered is general and can be implemented in a large number of systems, with the USC regime having been demonstrated experimentally for more than a decade~\cite{anappara2009signatures} ($\lambda/\omega > 0.1$). Since then, USC has been achieved in several systems. Recently, some experiments managed to reach the DSC, notably in Ref~\cite{yoshihara2017superconducting} in circuit quantum electrodynamics setup implementing the QRM, for the highest reported strength of light-matter interaction of $\lambda/\omega \approx 1.34$ with temperature of  approximately 45 mK ($k_{B}T/\hbar\omega \approx 0.16$), where thermal entanglement was also confirmed.-

\section{Acknowledgements}

H.-G. X. and J. J. are supported by National Natural Science Foundation of China under Grant No. 11975064. V.M. thanks the Postdoctoral Science Foundation of China (Grant No. 2022T150098)). G.X. acknowledges the support of NSF of China under Grants No. 12174346.
\newpage
\bibliographystyle{unsrt}
\bibliography{bibliography.bib}

\end{document}